\documentclass{PoS}

\usepackage{amsmath}
\usepackage{amsfonts}
\usepackage{amssymb}
\usepackage{caption}
\usepackage{float}
\usepackage{graphicx}
\usepackage{slashed}
\usepackage{xfrac}

\newcommand{\p}{\vec{p}}
\newcommand{\pp}{\vec{p}\:\!'}
\newcommand{\q}{\vec{q}}
\newcommand{\x}{\vec{x}}
\newcommand{\cG}{G}

\newcommand{\chib}{\overline{\chi}}
\newcommand{\phib}{\overline{\phi}}

\newcommand{\CSSM}{Special Research Centre for the Subatomic Structure
  of Matter (CSSM),\\School of Chemistry and Physics, University of
  Adelaide, South Australia 5005, Australia} 

\newcommand{\NCI}{National Computational Infrastructure
  (NCI),\\Australian National University, Australian Capital Territory
  0200, Australia} 

\newcommand{\CSIRO}{Digital Productivity Flagship, CSIRO,\\
	College Road, Sandy Bay, TAS 7005, Australia}

\title{Electromagnetic matrix elements for negative parity nucleons}

\ShortTitle{Electromagnetic matrix elements for negative Parity nucleons}

\author{\speaker{Benjamin Owen}%
         \thanks{We thank PACS-CS Collaboration for making their 2+1 flavor configurations available and acknowledge the
ongoing support of the ILDG. This research was undertaken with the assistance of resources at the NCI National Facility
in Canberra, Australia, and the iVEC facilities at Murdoch University (iVEC@Murdoch) and the University of Western
Australia (iVEC@UWA). These resources are provided through the National Computational Merit Allocation Scheme
and the University of Adelaide Partner Share supported by the Australian Government. We also acknowledge eResearch
SA for their support of our supercomputers. This research is supported by the Australian Research Council.}\\
        \CSSM \\
        E-mail: \email{benjamin.owen@adelaide.edu.au}}

\author{Waseem Kamleh, Derek Leinweber\\
        \CSSM}

\author{Selim Mahbub\\
				\CSIRO}

\author{Benjamin Menadue\\
				\CSSM \\
				\NCI}

\abstract{Here we present preliminary results for the evaluation of
  the electromagnetic form factors for the lowest-lying
  negative-parity, spin-$\sfrac{1}{2}$ nucleons, namely the
  $S_{11}(1535)$ and $S_{11}(1650)$, through the use of the
  variational method.  We find that the characteristics of the
  electric form factor, $G_{E}$, are similar between these states,
  however significant differences are observed between the
  quark-sector contributions to the magnetic form factor, $G_{M}$.
  Within simple constituent quark models, these states are understood
  to be admixtures of $s=\sfrac{1}{2}$ and $s=\sfrac{3}{2}$ states
  coupled to orbital angular momentum $\ell = 1$.  Our results reveal
  a qualitative difference in the manner in which the
  singly-represented quark sector contributes to these baryon magnetic
  form factors.  }

\FullConference{The 32nd International Symposium on Lattice Field Theory\\
                 23-28 June, 2014\\
                 Columbia University New York, NY}

\begin{document}

\section{Introduction}

Over the past decade there has been significant experimental interest
in mapping out the excited nucleon spectrum and understanding the
underlying dynamics and structure of these states. Such data provides
an excellent opportunity to connect experiment with theoretical
expectations to gain further insight into hadronic excitations. The
success of using variational techniques to explore the hadron spectrum
from Lattice QCD has shown that these methods can be utilised for
calculations of both ground state \cite{Owen:2012:PLB} and excited
state \cite{Owen:2012:PoS,Owen:2013} matrix elements. In this work we
perform an evaluation of the electromagnetic form factors of the
lowest-lying spin-$\sfrac{1}{2}$ negative-parity nucleon states.

\section{Accessing negative parity states}

The simplest approach to evaluating the correlators relevant to
accessing negative-parity nucleons \cite{Lee:1998cx} is to use the standard
nucleon interpolator coupled with an additional $\gamma_5$ matrix in
order to change its parity transformation properties, $\chi_{p}(x) \rightarrow
\chi^{-}_{p}(x) = \gamma_5 \chi_{p}(x)$. One then evaluates the
two-point correlator in the standard fashion and projects out the
state via the standard projection operator $\Gamma = \left(
\frac{\gamma_0 + I}{2} \right)$
\begin{align}
	\cG_{-}(\p, t; \Gamma) &= \sum_{\x} e^{-i \p \cdot \x}
        \mathrm{tr} \left[ \Gamma \, \langle \, \Omega \, | \,
          \chi^{-}_{p}(x)\, \chib^{-}_{p}(0) \, | \, \Omega \rangle
          \right] \\ 
												 &=
        - \sum_{\x} e^{-i \p \cdot \x} \mathrm{tr} \left[ \Gamma \,
          \langle \, \Omega \, | \, \gamma_5 \chi_{p}(x) \, \chib_{p}(0)
          \gamma_5 \, | \, \Omega \rangle \right]	\, . 
\end{align}
However using the cyclicity of the trace, one could instead access the
relevant contributions for negative parity states from the correlator
evaluated with the positive parity operator if one instead uses the
modified projector
\begin{equation}
	\Gamma^{-} = - \gamma_5 \Gamma \gamma_5 = \left( \frac{\gamma_0 - I}{2} \right) \, .
\end{equation}
Such an approach has long been established as the optimal method for
studying negative parity states.  However, we outline this in detail
here as the arguments carry over naturally to the evaluation of
three-point correlation functions.  Again, one could evaluate the
three-point correlator for a negative parity nucleon through
\begin{align}
	\cG^{\mu}_{-}(\pp, \p; t_2, t_1; \Gamma') &= \;\;\;\;
        \sum_{\x_2, \x_1} e^{-i \pp \cdot \x_2} e^{+ i (\pp - \p)
          \cdot \x_1} \mathrm{tr} \left[ \Gamma' \, \langle \, \Omega
          \, | \, \chi^{-}_{p}(x_2) j^{\mu}(x_1) \chib^{-}_{p}(0) \, |
          \, \Omega \, \rangle \right] \\ 
																			&=
        - \sum_{\x_2, \x_1} e^{-i \pp \cdot \x_2} e^{+ i (\pp - \p)
          \cdot \x_1} \mathrm{tr} \left[ \Gamma' \, \langle \, \Omega
          \, | \, \gamma_5 \chi_{p}(x_2) j^{\mu}(x_1) \chib_{p}(0)
          \gamma_5 \, | \, \Omega \, \rangle \right] \, . 
\end{align}
However, we can again access the necessary terms by evaluating
correlators with the positive parity operators and projecting with the
modified projector, $\left( \Gamma' \right)^{-} = - \gamma_5 \Gamma'
\gamma_5$.

\section{Variational Methods for matrix element determination}

The goal of the this approach is to produce a set of operators
$\phi^{\alpha}$ that satisfy
\begin{equation}
\label{diag}
\langle \Omega | \phi^{\alpha} | \beta, p, s \rangle = \delta^{\alpha
  \beta} \, .
\end{equation}
This is realised by taking an existing basis of operators $\lbrace
\chi_i \rbrace$ and constructing the desired operators as linear
superpositions
\begin{equation}
\phi^{\alpha}(x) = \sum_{i} v^{\alpha}_{i}\,  \chi_{i}(x) \:
, \hspace{20pt} \phib\,{}^{\alpha}(x) = \sum_{j} \chib_{j}(x)\,
u^{\alpha}_{j} \, . 
\end{equation}
Starting from the matrix of cross correlators
\begin{equation}
\cG_{ij}(\p, t; \, \Gamma) = \sum_{\x} e^{-i \p \cdot \x} \mathrm{tr}
\left[ \Gamma \langle \Omega | \chi_i(x) \chib_j(0) | \Omega \rangle
  \right] \, , 
\end{equation}
and noting $\cG_{ij}(\p, t; \Gamma)\, u^{\alpha}_{j}$ provides a
recurrence relation with time dependence $e^{-E_{\alpha} t}$, one can
show that the necessary vectors $v^{\alpha}_{i}$ and $u^{\alpha}_{j}$
are the eigenvectors of the generalised eigenvalue equations 
\begin{subequations}
	\begin{align}
	v^{\alpha}_{i} \: \cG_{ij}(\p, t_0 + \Delta t; \, \Gamma) \hspace{13pt} &
        = e^{-E_{\alpha} \Delta t} \: v^{\alpha}_{i} \: \cG_{ij}(\p, t_0; \, \Gamma) \, , \\
	\hspace{13pt} \cG_{ij}(\p, t_0 + \Delta t; \, \Gamma) \: u^{\alpha}_{j} & = e^{-E_{\alpha} \Delta t} \: \hspace{13pt} \cG_{ij}(\p, t_0; \, \Gamma) \: u^{\alpha}_{j} \, .
	\end{align}
\end{subequations}
It is worth noting that these equations are evaluated for a given
3-momentum $\p$ and projection operator $\Gamma$ and so the
corresponding operators satisfy Eq.~\eqref{diag} for this momentum and
parity only. One can obtain the correlator for the state $\alpha$ by
projecting with the corresponding eigenvectors
\begin{equation}
\cG^{\alpha}(\p, t; \Gamma) \equiv v^{\alpha}_{i}(\p) \: \cG_{ij}(\p, t; \Gamma) \: u^{\alpha}_{j}(\p) \, ,
\end{equation}
from which the desired quantities are extracted in the standard
way. To access the corresponding three-point correlator, it is a
simple matter of applying the relevant eigenvectors to the
corresponding three-point function, where care is taken to ensure that
the projection is done with the correct momenta for source and sink
\begin{equation}
	\cG^{\alpha}(\pp, \p, t_2, t_1; \, \Gamma') \equiv v^{\alpha}_{i}(\pp) \: \cG_{ij}(\pp, \p, t_2, t_1; \, \Gamma') \: u^{\alpha}_{j}(\p) \, .
\end{equation}
From the projected two and three-point functions one then continues on
in the standard way by constructing a suitable ratio to isolate the
desired matrix element. Here we choose to use the ratio as defined in
Ref.~\cite{Hedditch:2007}. Using the modified projectors outlined in
the previous section, we arrive at the ratio used in the determination
of the form factors
\begin{equation}
	R^{\alpha}_{-}(\pp, \p; \Gamma', \Gamma) = \sqrt{ \frac{\langle \cG^{\alpha}(\pp, \p, t_2, t_1; \left(\Gamma'\right)^{-}) \rangle \, \langle \cG^{\alpha}(\p, \pp, t_2, t_1; \left(\Gamma'\right)^{-}) \rangle}{\langle \cG^{\alpha}(\pp, t_2; \Gamma^{-}) \rangle \, \langle \cG^{\alpha}(\p, t_2; \Gamma^{-}) \rangle} } \, .
\end{equation}

\section{Negative Parity Baryon Form Factors}

Here we consider the electromagnetic form factors for a negative
parity, spin-$\sfrac{1}{2}$ baryon. To make the connection with the
familiar positive parity case, we note that negative-parity baryon spinors
can be defined relative to positive parity spinors by again
multiplying by a $\gamma_5$ matrix and attributing the
odd-parity baryon mass to considerations of $p$ 
\begin{equation}
	u(p, s) \rightarrow u_{-}(p, s) = \gamma_5\, u(p, s) \, .
\end{equation}
One can then show through the vertex decomposition presented in
Refs.~\cite{Devenish:1975jd,Aznauryan:2008us} that it is possible to
write the $+ \rightarrow +$ and $- \rightarrow -$ transitions, of
which the elastic processes are a special case, in a common form. With
this understanding it follows that the matrix element can be expressed
as\footnote{We note that this result could also be obtained through
  the freedom to choose an intrinsic parity for the baryon spinor.
  However, this discussion is valuable in the consideration of
  parity-changing electromagnetic transitions.}
\begin{equation}
	\langle \, N^{-}, p', s' \, | \, j^{\mu}(0) \, | \, N^{-}, p, s \, \rangle = \left( \frac{M^2}{E_{p} E_{p'}} \right)^{\sfrac{1}{2}} u(p', s') \left( \gamma^{\mu} F_{1}(Q^2) + i \frac{\sigma^{\mu\nu}q_{\nu}}{2M} F_{2}(Q^2) \right) u(p,s) \, .
\end{equation}
These are in turn related to the Sachs Electric and Magnetic form factors
\begin{align}
	G_{E}(Q^2) &= F_{1}(Q^2) - \frac{Q^2}{(2M)^2} F_{2}(Q^2) \, ,\\
	G_{M}(Q^2) &= F_{1}(Q^2) + F_{2}(Q^2) \, .
\end{align}
To isolate these form factors, we follow the approach outlined in
Refs.~\cite{Leinweber:1990,Boinepalli:2006}. Having projected out the
correlators relevant for the state $| \alpha \rangle$ and formed the
necessary ratio, we choose the incoming state to be at rest and so
extract $G_E$ and $G_M$ through the following terms
\begin{equation}
	G_E(Q^2) = \overline{R}^{4}_{-}(\q, 0; \Gamma_4, \Gamma_4) \, , \quad G_M(Q^2) = \frac{(E_{q}+M)}{|\q|} \overline{R}^{3}_{-}(\q, 0; \Gamma_2, \Gamma_4) \, ,
\end{equation}
where $\overline{R}$ is the reduced ratio
\begin{equation}
	\overline{R}^{\mu}_{-}(\pp,\p; \Gamma) = \left[ \frac{2 E_p}{E_p + M} \right]^{\sfrac{1}{2}} \left[ \frac{2 E_{p'}}{E_{p'} + M} \right]^{\sfrac{1}{2}} R^{\mu}_{-}(\pp,\p; \Gamma) \, .
\end{equation}

\section{Calculation Details}

The states of interest, the $S_{11}(1535)$ and the $S_{11}(1650)$,
have been isolated in a previous CSSM study \cite{Mahbub:2012} and so
we shall use the same operator basis and parameters in our variational
analysis. To form our operator basis we use local nucleon operators
\begin{equation}
	\chi_1(x) = \epsilon^{a b c} \left( u^{T a}(x)\, C \gamma_5\,
        d^b(x) \right) u^c(x) \, , \quad \chi_2(x) = \epsilon^{a b c}
        \left( u^{T a}(x)\, C \, d^b(x) \right) \gamma_5 u^c(x) \, , 
\end{equation}
coupled with varying levels of gauge-invariant Gaussian smearing
applied to both fermion source and sink. In particular, we use 16, 35,
100 and 200 sweeps of smearing applied to the spatial dimensions only,
with a smearing fraction $\alpha = 0.7$. This allows for the
construction of an 8 $\times$ 8 correlation matrix. For the
variational parameters we use $t_0 = 18$ and $\Delta t = 2$ relative
to the quark source at $t=16$. The calculation is performed on the
PACS-CS 2 + 1 flavour dynamical gauge-field configurations
\cite{Aoki:2008} made available through the ILDG \cite{Beckett:2011}.
These configurations use an ${\cal O}(a)$-improved Wilson-Clover
fermion action and Iwasaki gauge-action, with $\beta = 1.90$ resulting
in a lattice spacing $a = 0.0907$~fm. The lattices have dimension
$32^3$ $\times$ $64$ giving rise to a spatial box of length $L =
2.9$~fm. We have access to five light quark masses, with the strange
quark mass held fixed. The resulting pion masses range from 702~MeV
right down to 156~MeV. The resulting spectrum is presented in Fig.~1.
We observe two low-lying eigenstates with small mass difference,
consistent with the experimentally observed masses for the
$S_{11}(1535)$ and $S_{11}(1650)$. It is these states whose form
factors we shall examine herein.

\begin{figure}[t]
\centering
\hspace*{-30pt}
\includegraphics[width=0.75\textwidth]{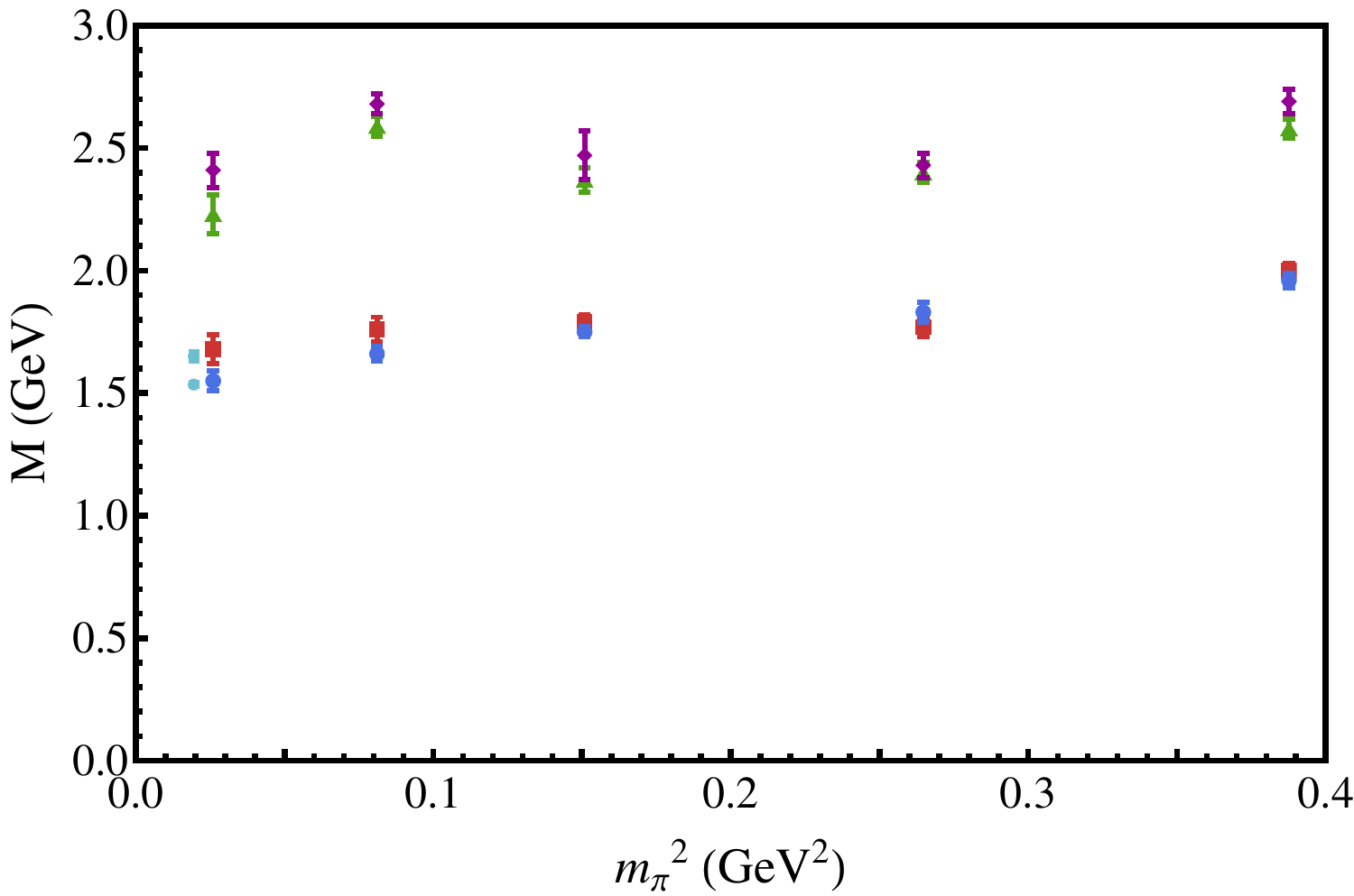}
\caption{The four lowest lying-states observed in our $\sfrac{1}{2}^-$
  nucleon spectrum obtained via an 8 $\times$ 8 correlation matrix
  formed from smeared $\chi_1$ and $\chi_2$ interpolators.  The light
  blue data points correspond to the PDG values \cite{Agashe:2014} for
  the negative parity nucleon states with 3-star determination or
  higher.}
\end{figure}

For the SST inversion we choose to use the fixed current method with a
conserved-vector current inserted at $t_S = 21$. For our error
analysis we use a second-order single-elimination jackknife method
where the $\chi^2_{\textrm{dof}}$ is obtained via a covariance matrix
analysis. For the form factors we consider all but the lightest quark
mass. The eigenstate projected correlators are fitted to a single
state ansatz. By studying the regions where $\log(G)$ behaves
linearly we ensure that the correlator is dominated by a single energy
eigenstate. Further discussion can be found in
Refs.~\cite{Owen:2013,Mahbub:2013}. 

\section{Results}

\begin{figure}[t]
\centering
\hspace*{-5pt}
\includegraphics[width=0.49\textwidth]{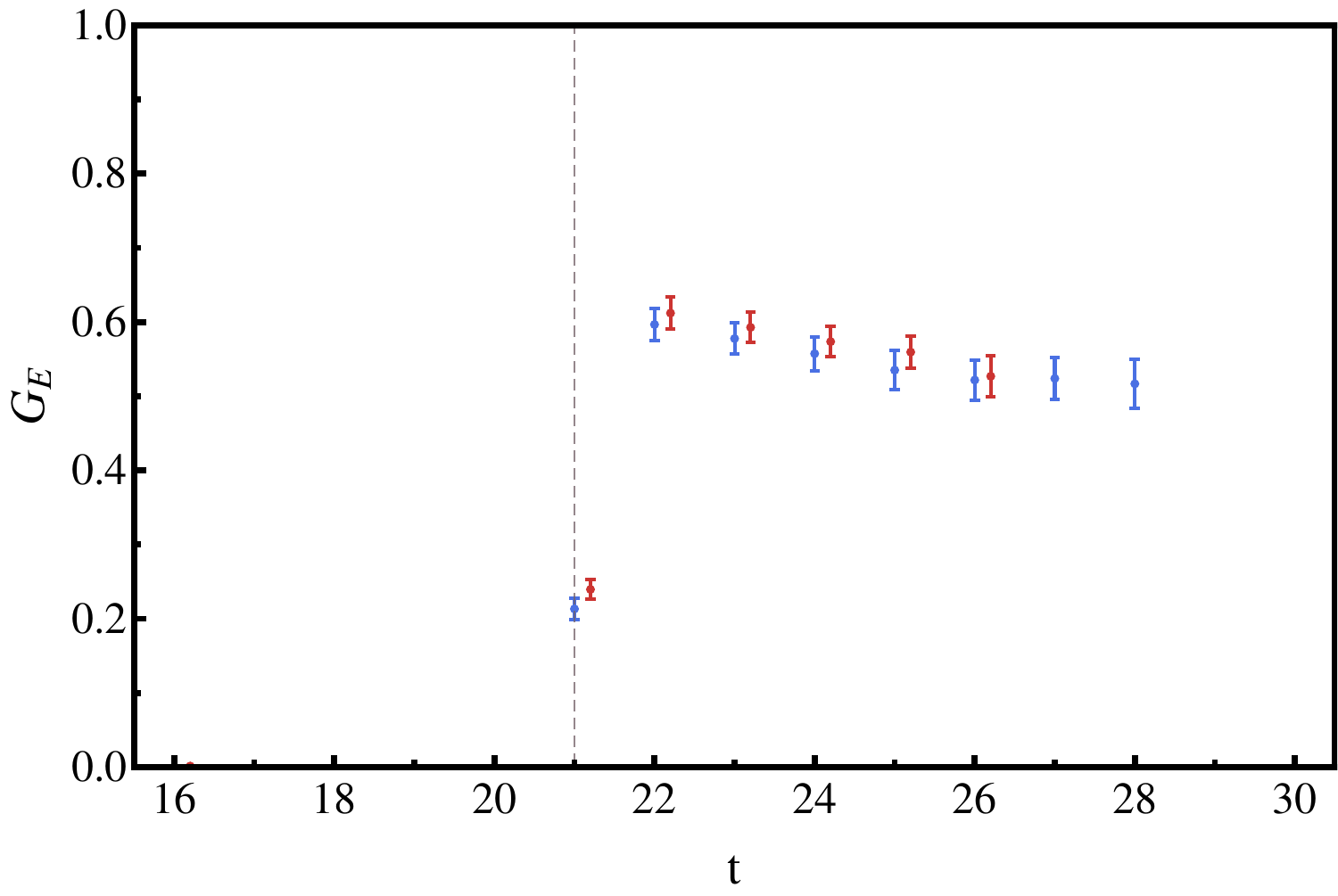}
\includegraphics[width=0.49\textwidth]{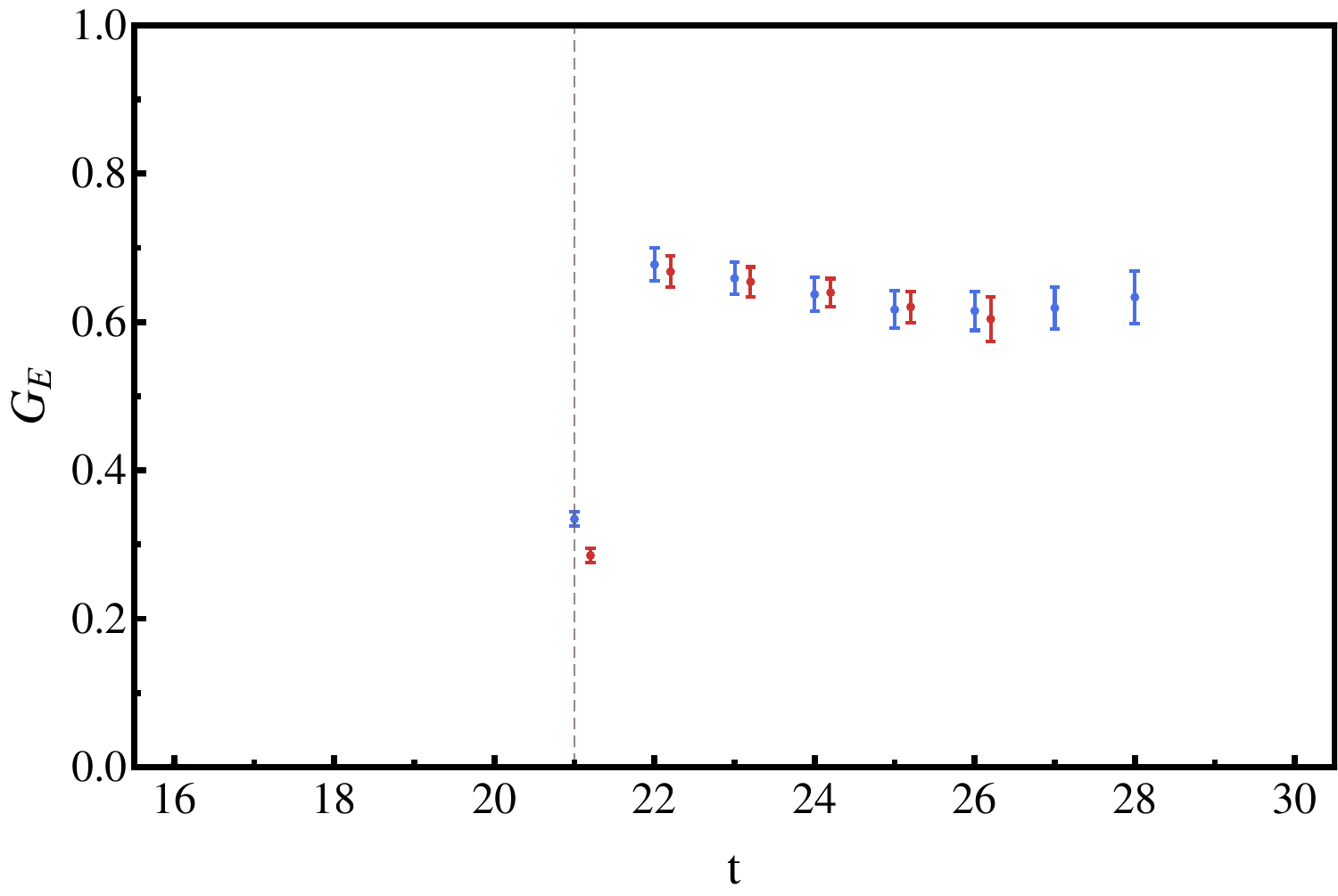}
\caption{The doubly-represented (left) and singly-represented (right)
  quark sector contributions to the the Sachs electric form factor
  $G_E$ for $m_{\pi} = 0.570$~GeV.  Results are provided for single
  quarks of unit charge.  The blue data points correspond to the
  $S_{11}(1535)$ while the red data points to the $S_{11}(1650)$.}
%
\vspace{8pt}
\hspace*{-5pt}
\includegraphics[width=0.49\textwidth]{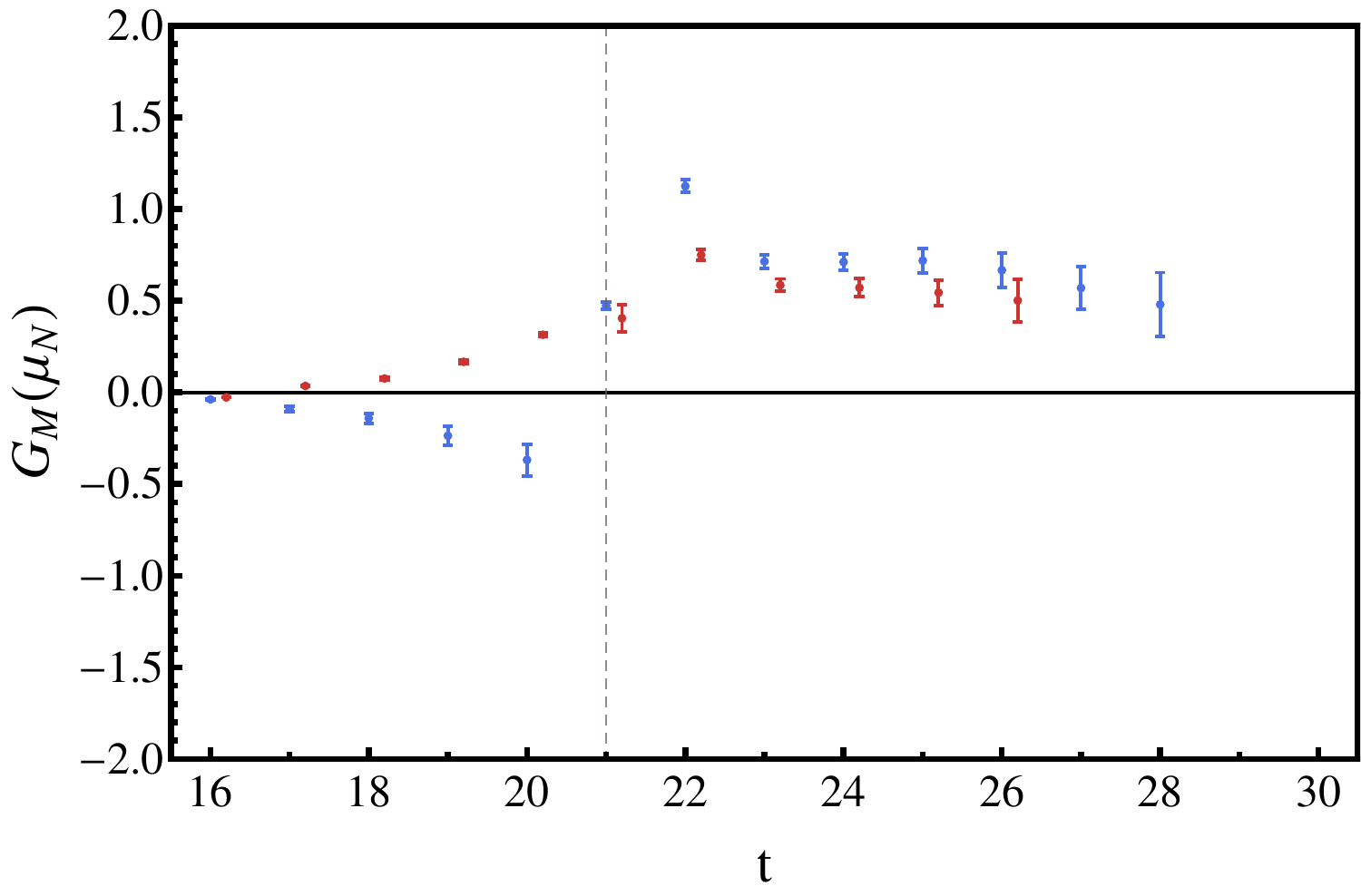}
\includegraphics[width=0.49\textwidth]{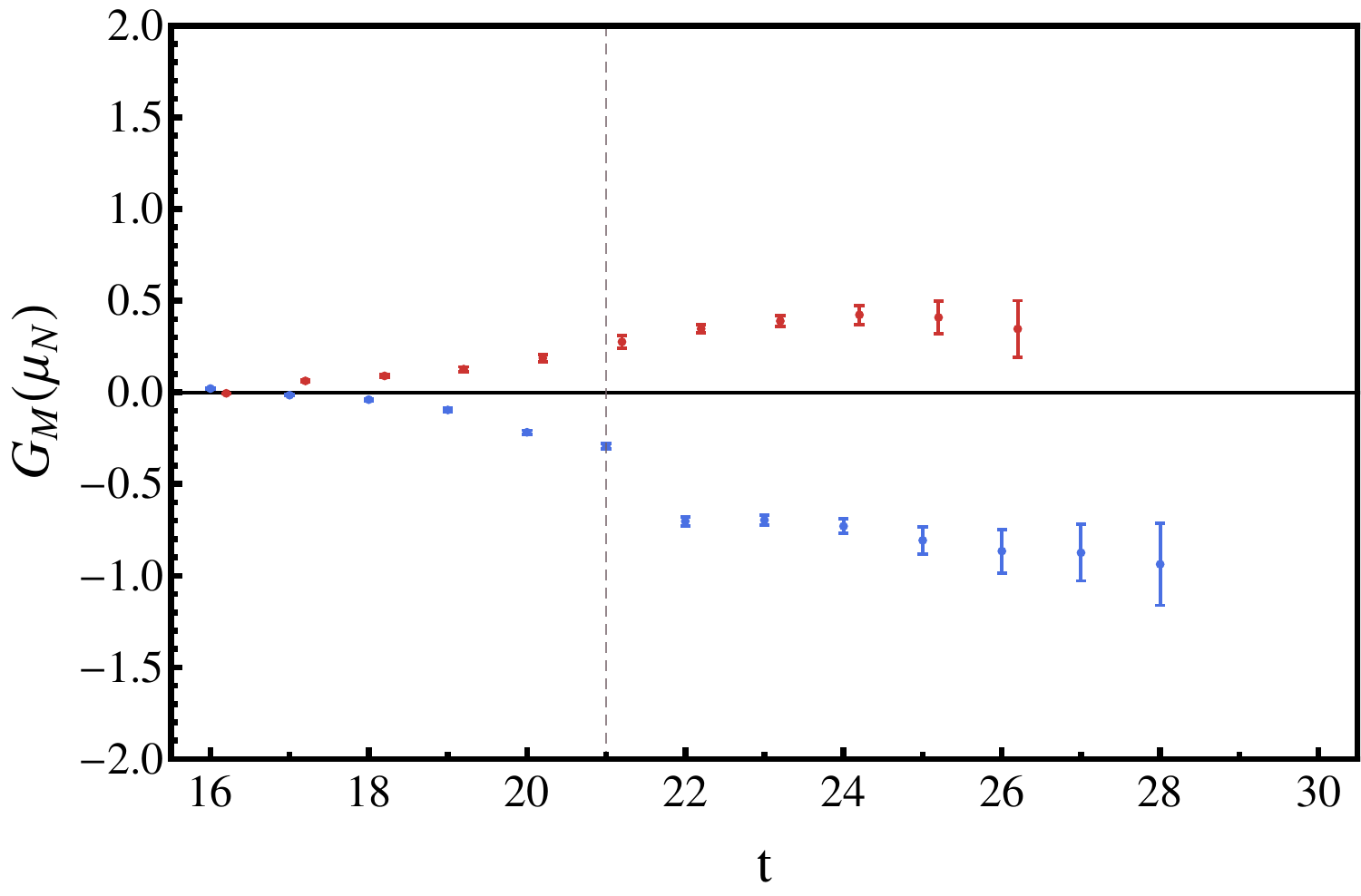}
\caption{The doubly-represented (left) and singly-represented (right)
  quark sector contributions to the the Sachs magnetic form factor
  $G_M$ for $m_{\pi} = 0.570$~GeV.  Results are provided for single
  quarks of unit charge.  Again, the blue data points correspond to
  the $S_{11}(1535)$ while the red data points to the $S_{11}(1650)$.}
\end{figure}

In Figs.~2 and 3 we present the quark-sector results for the electric
and magnetic form factors respectively.  We present results for single
quarks of unit charge.  The data is presented at a single quark mass
corresponding to $m_\pi = 0.570$ GeV.  However, all masses considered
display behaviour consistent with that in Fig.~2 and 3. The colours
match up with the states presented in Fig.~1, with the blue identified
as the $S_{11}(1535)$ and the red as the $S_{11}(1650)$.

We note that due to the similar masses between these two states, we
are probing each state at essentially the same value of $Q^2$. For the
electric form factor, we find that both states take on very similar
values in both quark sectors, with the doubly-represented quark-sector
form factor slightly smaller than the singly-represented
sector. Examining the magnetic quark sector we see distinctly
different behaviour between these two states. Though the
doubly-represented quark sector is similar, we find that the
singly-represented quark sector in the $S_{11}(1650)$ is positive
(same sign as the doubly-represented sector) while the corresponding
contribution in the $S_{11}(1535)$ is negative.  Furthermore, the
magnitude of the singly-represented quark contribution is somewhat
larger in the $S_{11}(1535)$ than it is in the $S_{11}(1650)$. 


\section{Conclusions}

Herein we have presented the first lattice QCD calculation of the
electromagnetic form factors of the two lowest-lying
spin-$\sfrac{1}{2}$ negative-parity nucleons.  Using variational
techniques we are able to disentangle the relevant matrix element for
these two states, allowing us to probe their underlying structure.

Both states display very similar values in their electric form factor.
However, comparison of the magnetic form factor highlights distinctly
different behaviour in the quark sector contributions.  Such behavior
is anticipated in simple quark models due to differences in the
underlying space-spin-flavour symmetry construction of these states.

The difference in the sign between the doubly-represented and
singly-represented quark contributions in the $S_{11}(1535)$ and the
sign symmetry of quark sector contributions in the $S_{11}(1650)$ will
give rise to significantly different baryon form factors.  It will be
interesting to compare these with phenomenological estimates and gain
insight into the mechanisms of QCD giving rise to these observations.
Future work will generalise these techniques to examine the
corresponding helicity amplitudes for these states, central to the
radiative transitions measured in experimental programs.

\end{document}